\documentclass[runningheads]{llncs}
\newcommand{\F}{\mathbb{F}}
\usepackage{comment}
\usepackage{multicol}
\usepackage{array}
\usepackage{amssymb}
\usepackage{amsmath}
\usepackage{graphicx}
\usepackage{multirow}
\usepackage{tikz}
\usepackage{enumerate}
\usepackage{booktabs}
\begin{document}
\title{Premium Access to Convolutional Neural Networks}
%
%
\author{Julien Bringer\inst{1} and Herv\'e Chabanne\inst{2,3} and Linda Guiga\inst{2,3}}

\institute{Kallistech \\
\email{julien@kallistech.com} 
\and 
Idemia\\
\email{firstname.lastname@idemia.com}
\and
T\'el\'ecom Paris
}
\maketitle              
\begin{abstract}
Neural Networks (NNs) are today used for all our daily tasks; for instance, in mobile phones. We here want to show how to restrict their access to privileged users. Our solution relies on a degraded implementation which can be corrected thanks to a PIN. We explain how to select a few parameters in an NN so as to maximize the gap in the accuracy between the premium and the degraded modes. We report experiments on an implementation of our proposal on a deep NN to prove its practicability.

\end{abstract}
\section{Introduction}
Today, mobile phones are daily used for various tasks. In parallel, many of their applications rely on  Neural Networks (NNs) to perform different classification tasks. We here want to investigate how to restrict their use to a set of privileged users, who have access to a premium mode. Here, the premium mode is defined per user by training his NN for his personal use.  \par 
The first idea behind our solution comes from \cite{hacking}. \cite{hacking} describes how, in 2016, mobile RSA’s SecurID and Vasco DIGIPASS Software Tokens can be hacked despite relying on different defense mechanisms implemented to thwart reverse engineering processes. Its conclusion is that in such a hostile environment, quoting: ``The best defense against the attacks shown in this paper is securing the mobile token with a PIN''.

As our protection relies on low-entropy PINs, we can not let hackers perform brute-force attacks at ease.
We are going to implement a degraded classification task by default for wrong PINs vs an optimal one for the privileged users; rather than implementing  simple work/no-work modes for the application. Switching from a degraded mode to the premium optimal one is done by modifying some parameters of the NN. A particular emphasis is given to the way we store these critical parameters in the mobile phone (see Sec. \ref{method} for details).
\par 
Our second idea relies on some specific layers: the convolutional ones. These are found in Convolutional Neural Networks (CNNs), which are the most used NNs for image processing, for instance. We select some parameters in a given convolutional layer as the ones enabling us to move from degraded modes to the optimal one (see Sec. \ref{Select}).\par

Going a step further, we exploit the fact that we are dealing with NNs. During a training phase with a dedicated database, the parameters of an NN are optimized. We are going to introduce a strategy which uses the fact that an attacker who does not have full access to testing facilities will not have the capacity to find optimal parameters through retraining. \par

To sum up, each privileged user comes with an NN trained for him and a PIN enabling him to reach the premium mode of his NN. \par 
We end this introduction with a description of some Related Works. In Sec. \ref{NNs}, we recall some facts about NNs. We describe our proposal to store optimal parameters  in Sec. \ref{method}. To illustrate its practicability, we report in Sec. \ref{Experiments} our experiments on an NN based on ResNet18 \cite{resnet} which is typical of deep learning.  Sec. \ref{Conclusion} concludes.
\subsection{Related Works} \label{RelatedWorks}
Android offers a multi-layer security strategy. \cite{DBLP:journals/corr/abs-1904-05572} describes this security model (see also \cite{android}).
Moreover, one may add ad-hoc protections for obfuscating the code making it harder to reverse-engineer \cite{eurocrypt-2016-28982}.
There is an on-going cat-and-mouse game between hackers and developers. However, it seems to us that developers may have a hard time whenever the full access to the code is available to hackers. For instance, for white-box cryptography, where the code for encrypting with DES or AES symmetric algorithms are given to attackers, all academic proposals have been broken so far \cite{fse-2016-29796}. Moreover, while complementary to our proposal, anti-reverse engineering techniques tend to inflate the size of the code a lot. For instance, for DES or AES, 
there may be a multiplication by 16k of the code size, from less than 1KB for an unprotected implementation to more than 16MB. As NNs are initially big, for instance the one we are considering here, such an expansion cannot be handled in an embedded environment. \par 

Our approach is different. We want to force the hacker to measure the performances of the deployed NN in mobile phones without having the possibility to rely on a dedicated database. We here study an example of facial recognition (see \cite{DBLP:journals/corr/abs-1804-06655} for a survey  of this domain). 
We think it is relevant, as a typical deep learning task.  \par

For a secure hardware-based implementation of NNs on mobile devices, see \cite{tubiblio117658}. \par

For a comprehensive study on the deployment of deep learning Android apps, see \cite{DBLP:conf/www/XuLLLLL19}.

\section{Background} \label{background}
\subsection{Convolutional Neural Networks} \label{NNs}
Today, Convolutional Neural Networks (CNNs) are used for making predictions in various fields of  application ranging from image processing \cite{processing}, to classification \cite{classification,vgg} and segmentation \cite{segmentation}.

They are composed of several layers: 
\begin{itemize}
\item Convolutional layers compute a convolution between one -- or several -- filter $F$ and the input, as follows:
$$
O_{i, j} = \sum_{k=1}^{h}\sum_{l=1}^{w} X_{i+k, j+l}\cdot F_{k, l}
$$
where $O$ is the output of the convolution. A convolution can be seen in Fig. \ref{fig:conv}. 
\begin{figure}
\centering
\includegraphics[scale=0.2]{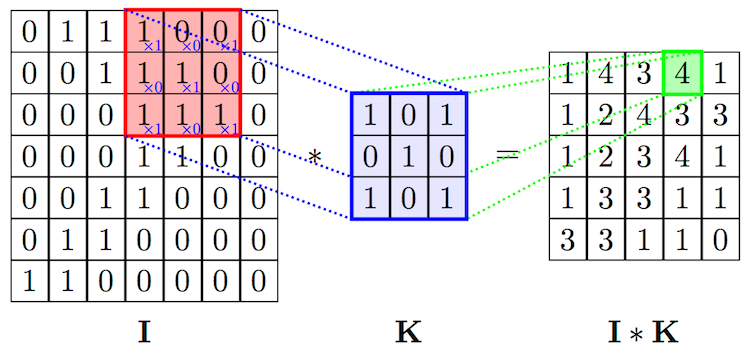} 
\caption{Convolution between an input $I$ and a filter $K$ }
\label{fig:conv}
\end{figure}
\par 
The elements of the filter are the weights of the layer and will be designated either by `weights' or by `parameters' in the rest of the paper. 
\item Other layer types include fully connected layers 
-- through weights -- to all the elements from the previous layer. 
\item A nonlinear function 
is applied at the end of each layer. The most popular one is the ReLU (Rectified Linear Unit) function defined as the max of a value and zero. It is used to activate -- or deactivate -- elements of the layer.

\item Finally, pooling layers are usually present between other layers in order to reduce the dimensionality of the input.
\end{itemize}
The input of each layer consists of different channels. For instance, in image processing, the input of the model is usually divided in three channels corresponding to the RGB colors. \par
The weights -- and other parameters -- of a CNN are trained over several epochs -- i.e. runs on a training data set -- so as to reach a value guaranteeing the best possible prediction accuracy. Given their large number of parameters, and the necessity of high accuracy nowadays, some NNs take days -- or even months -- to train. 
\par
Several techniques are added over the years to make training more efficient. One of these consists in adding a Batch Normalization layer to improve the training phase. In 2015, the authors of \cite{batchnormIoffe} discover this type of layer whose purpose is to make training faster, more efficient and more stable. The layer normalizes its input. Thus, given an element $x_{i, j}$ in a batch $B$ of its input, the layer computes:
\begin{equation} \label{batchnorm}
\tilde{x}_{i, j} = \gamma \frac{x_{i, j} - \mu_B}{\mathbb{V}_B} + \beta
\end{equation}
where $\gamma$ and $\beta$ are parameters optimized during the training phase, and $\mathbb{V}_B$ and $\mu_B$ are the considered batch's variance and expected value respectively. \par 

\subsection{ResNet18}
\begin{figure} 
    \centering
    \includegraphics[scale=0.3]{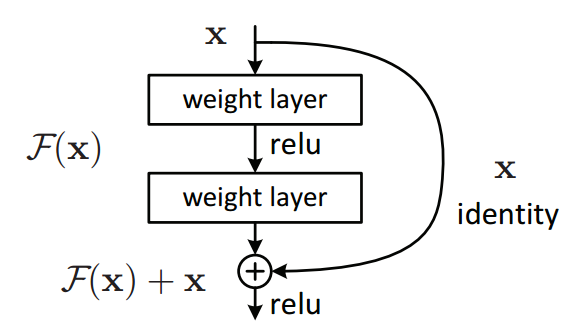}
    \caption{Residual Block in ResNet}
    \label{shortcut}
\end{figure}

At first glance, one could imagine that the more layers an NN contains, the better its accuracy will be, once fully trained. However, a known problem occurs for deep neural networks during training. In 2016, the authors of \cite{resnet} discover ResNet as a way to better train deeper NNs, without having to deal with it. This is achieved thanks to residual blocks corresponding to ``identity shortcuts", described in Fig. \ref{shortcut}. More generally, the architecture of Residual Neural Networks introduces these skipping connections. The authors argue that thanks to the identity mapping, the training should be similar be it with or without the shortcut layer. This is why large ResNet architectures, containing sometimes up to 1,001 layers \cite{deepresnet}, are efficiently trained with a high accuracy. \par
\begin{figure}
    \centering
    \includegraphics[scale=0.3]{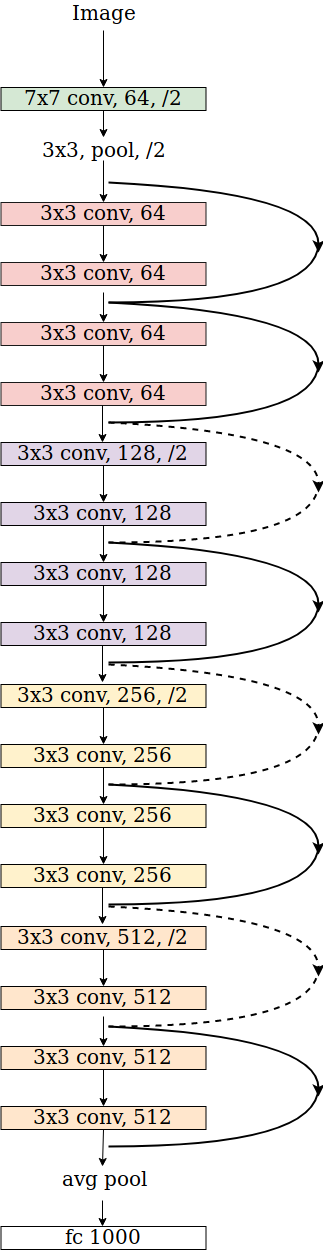}
    \caption{Architecture of ResNet18, where the dotted arrows indicate shortcuts increasing the dimension}
    \label{architecture}
\end{figure}

The model we consider here is based on a particular instance of ResNet: ResNet18. The latter is composed of 17 convolutional layers, a fully connected layer, a max pooling layer and a final global average pooling layer. The full architecture is described in Fig. \ref{architecture}. \par

\subsection{Selecting Optimal Parameters} \label{Select}
Our selection strategy aims at choosing only a few parameters -- called optimal -- following two criteria:
\begin{enumerate}
\item they must have an impact on the accuracy of the NN;
\item there must not be an obvious way for an attacker to find them other than by a try and guess method.
\end{enumerate}


Sec. \ref{Experiments} is devoted to detailing how this works on a particular instance of an NN. We will report how the fulfillment of the first criteria is achieved in this case. We also address the second criteria to penalize an attacker who has a limited access to the training database (see Sec. \ref{batchnormSec}).

 \subsubsection{Minimizing the Number of Parameters} \label{minparams}
The authors of \cite{oneneuron} show that some neurons have a higher impact on the model's prediction than others. Indeed, \cite{oneneuron} defines a neuron $i$'s sensitivity given an input $x$ as follows:
$$
\Delta(i, x) = argmin_{\delta} \{ |\delta| | f(x) \neq \tilde{f}_{\delta}^i (x)\}
$$
where $f$ is the original model and $\tilde{f}$ is a modified model where noise $\delta$ was added to the output of neuron $i$.

It corresponds to the minimal noise one needs to add to neuron $i$ for the classification to change. The authors of \cite{oneneuron} observe that a large number of neurons have a high sensitivity (small $\Delta$).

This result shows that it is possible to select few parameters to protect, and still prevent the attacker from getting a good accuracy. 

\subsubsection{When the Attacker Does Not Have Access to the Database} \label{staticdynamic}
\cite{Denil2013b} operates a distinction between static and dynamic parameters. We will also make such a distinction, but our definition of static parameters is slightly different from theirs. Let us define the following:
\begin{enumerate}
	\item We say that a parameter $w$ remains unchanged from one training epoch to the next if \\
	$w_{current\_layer} < r \cdot w_{previous\_layer}$ where $r = 10^{-2}$
	\item We denote \textit{static parameters} the parameters that have not changed over the last epoch. 
	\item \textit{Dynamic parameters} are the non-static parameters
\end{enumerate} 
The choice of $r$ in Point 1. comes from the fact that a slight change in a parameter does not lead to a noticeable drop in the accuracy. What interests us when studying the parameter fluctuation is the way the modifications influence the accuracy. Thus, $r$ is tuned so that the resulting evolution curves for the number of static parameters is representative of the evolution of the accuracy. I.e. $r$ is chosen so that when the accuracy changes less, the number of static parameters increases drastically. \par 
Static parameters are easier to obtain by the attacker through a shorter training. Moreover, dynamic parameters are the ones that change the accuracy over the last few epochs and bring it to its optimal value. For those two reasons, protecting the dynamic parameters seems to be a viable strategy in order to limit the number of parameters to protect.

\subsection{Per user training}
For a same NN architecture, training with different initialization parameters results in different weights for all layers (see \cite{init}). As our privileged users benefit from dedicated training, they do share the same NN architecture, but with different parameters. For our proposal, this means that we have to modify the optimal parameters for each of the privileged users' NN.
To the best of our knowledge, given a trained NN, there is no way to deduce the parameters computed through another training with a different initialization of the same NN. 
\section{Protecting Optimal Parameters} \label{method}

Here, we suppose we have a set of $n$ optimal parameters $\{o_1, \ldots, o_n\}$ that we want to keep secret.
These secrets have to be protected by a PIN, in such a way that an attacker cannot proceed to an exhaustive search among all PIN values. I.e. for all PIN values, our mechanism has to return legit outputs. \par


Let $\F$ denote a finite field with $2^l$ elements such that $2^l -1$ is a prime. For instance, $\F = \F_{2^{521}}$.

We want to keep $l$ small. This means that we want a small $n$ too. An example of doing that is given in the next section. 

\begin{lemma} \label{lemma}
\begin{enumerate}
\item The non-zero elements of $\F$ form a multiplicative group.
\item This group is cyclic.
\item In this group, all elements are generators except the unity. 
\end{enumerate}
\end{lemma}

Point 2. of the previous lemma means that all non-zero elements can be expressed as powers of a single element called a generator. \par

For a proof, see \cite{Lidl:1997}.

Denote $O=*|o_1|\ldots|o_n \in \F$
where $|$ stands for the concatenation and $*$ is a bitstring with no particular value which is introduced to fit the length of the finite field $\F$ elements. \par

Given a PIN value, we then compute \begin{equation} \label{eq} g = O^{1/\text{PIN}} \in \F \end{equation}
Note that PIN is always invertible modulo $p = 2^{l} - 1$ since $p$ is taken to be a prime. Moreover, $g$ always satisfies (\ref{eq}) according to Lemma \ref{lemma}. \par
We store the function $f: \pi \mapsto g^{\pi} \in \F$ in mobile phones, we have: $f(\pi) = O$ if and only if $\pi=$ PIN. \par 

Thanks to Point 3. of Lemma \ref{lemma}, $g$ is a generator of the multiplicative group of $\F$ and $f(\text{all the values between 1 and } 2^l-1) = \F \setminus \{0\}$, which implies that an attacker who tries all possible values of PIN will get all elements of $\F$. This way, she will have no clue on which one has been chosen for $O$.  

Note that the implementation of function $f$ does not have to be secured.

\section{Example of Application: Facial Recognition} \label{Experiments}

In this section,  we consider an adapted version of the ResNet18 \cite{resnet} model architecture to the task of facial recognition. 
We think that this is a relevant example, as it demonstrates the feasibility of our concept on an NN structure that is used in different applications. Moreover, relying on facial recognition facilitates experiments on large datasets and comparisons with large scale benchmarks. Our architecture extends ResNet18 and relies on 14 million parameters across 76 layers. Our goal is to extract at most around a hundred parameters.

For facial recognition, the performances are assessed thanks to the accuracy of the recognition. 
On the one hand, false positives might happen, allowing unauthorized individuals to be recognized. On the other hand, false negatives might be a nuisance to genuine users.
More precisely,  the error is measured as follows: given a maximal False Acceptance Rate (FAR) --  i.e. the probability of a malicious individual being authenticated, assess the False Rejection Rate (FRR) -- i.e. the probability of a genuine user being rejected. 

The accuracy of our ResNet18-based model on the Labeled Faces in the Wild (LFW) database \cite{LFW} in our proprietary setting is as follows (3):
\begin{itemize}
	\item For $FAR = 10^{-4}$, $FRR = 0.24$ \%
	\item For $FAR = 10^{-5}$, $FRR = 0.70$ \%
\end{itemize} 
In our case, we reach the best accuracy after 13 training epochs.

\subsection{Optimal Parameters for ResNet18} \label{OptimalParametersforResNet18}

Given the large number of parameters in our NN model, carefully selecting the parameters to protect allows us to limit the size of $\F$ (see Sec. \ref{method}). In the following section, we describe two main parameter selection strategies. In the first, we protect parameters from batch normalization layers, either by protecting all the parameters of one layer, or by using the method described in \ref{staticdynamic}. In the second, we describe a strategy to sample elements from a convolutional layer, as a way to limit the number of optimal parameters.  



\subsection{Batch Normalization}\label{batchnormSec}
\subsubsection{Generating Suboptimal Parameters} \par 
When selecting a set of parameters to protect in an NN, the first, most intuitive, strategy would be to protect a layer with few parameters. As explained in Sec. \ref{background}, batch normalization layers aim at normalizing the input. For this reason, each of the layer's parameters affect one whole input channel. The said parameters are therefore scarce and impactful. Thus, batch normalization layers are one obvious choice of layer to protect. More specifically, we will focus here on the $\gamma$ parameters mentioned in (\ref{batchnorm}). 
We randomize the parameters of a batch normalization layer in the middle of the architecture (38th layer out of 76). The layer contains 128 $\gamma$ parameters. 
\par
\begin{figure}
	\centering
	\includegraphics[scale=0.45]{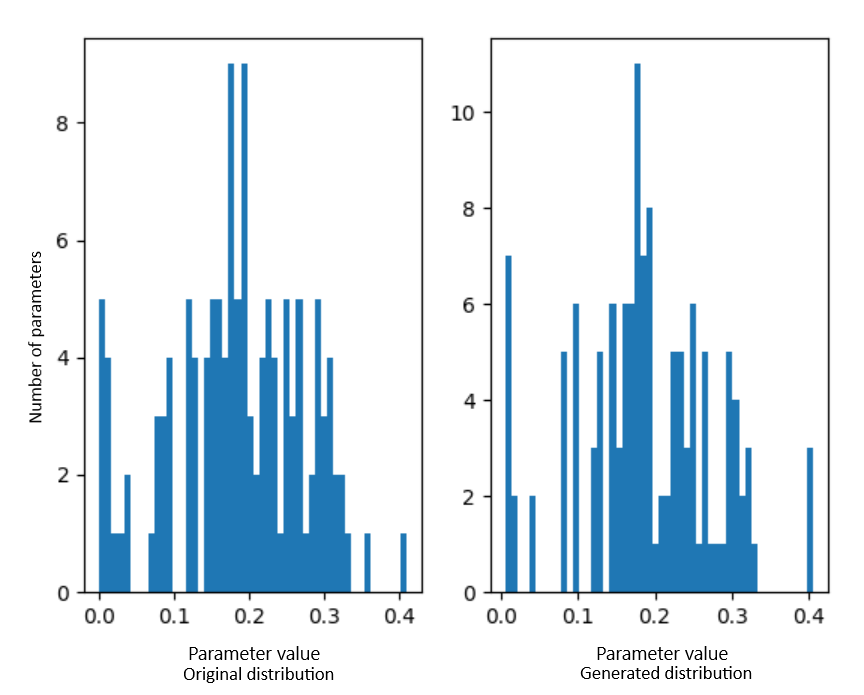} 
	\caption{Distribution of the $\gamma$ parameters in a batch normalization layer from the ResNet18 network, with the original distribution on the left and the generated distribution on the right}
	\label{gamma_dist}
\end{figure}

 
Even though the attacker does not have access to trained weights, observing the other batch normalization layers might enable them to spot erroneous settings if the random parameters selected do not reflect the usual distribution of $\gamma$ parameters. To prevent this, we compute the distribution of the chosen layer's $\gamma$ parameters and generate values following the same distribution (see Fig. \ref{gamma_dist}).

Once we have established the way suboptimal parameters have been generated, we can observe the associated drop in the accuracy and evaluate the security of our process for this strategy. \par 
When we change the selected layer's $\gamma$ parameters to random ones following the distribution of batch normalization layers, we get the following accuracy:
\begin{itemize}
	\item For $FAR = 10^{-4}$, $FRR = 0.32$ \%
	\item For $FAR = 10^{-5}$, $FRR = 0.96$ \%
\end{itemize}
Thus, the false rejection rate for $FAR = 10^{-4}$ has increased by $33\%$ and the rate for $FAR = 10^{-5}$ has increased by $37.1\%$ compared to the original model (see (3) for reference). This corresponds to the accuracy of the model after only 8 training epochs. 
Thus, modifying only one small layer over the 76 ones already results in a critical drop in the accuracy. Protecting the 128 $\gamma$ parameters of the batch normalization layer would therefore be enough to distinguish between premium and degraded accesses.\par 
The following section describes a second strategy. 


\subsubsection{Static VS Dynamic}
\begin{figure}
    \centering
    \includegraphics[scale=0.45]{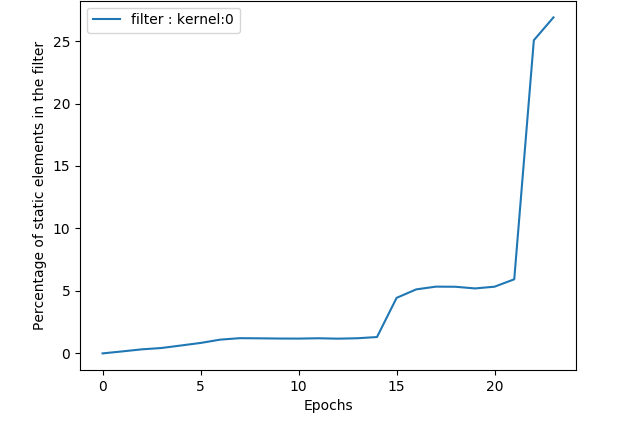}
    \caption{Percentage of static parameters in a convolutional filter with relation to the epoch.}
    \label{fig:stage2_conv1}
\end{figure}

\begin{figure}
    \centering
    \includegraphics[scale=0.45]{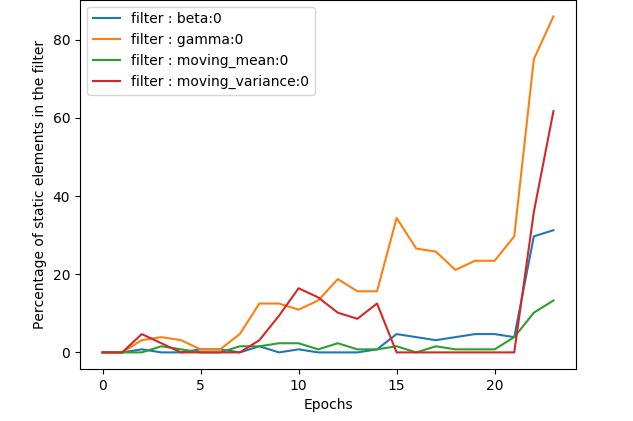}
    \caption{Percentage of static parameters for each parameter type in a batch normalization layer with relation to the epoch.}
    \label{fig:stage2_bn3}
\end{figure}

\begin{figure}[ht]
\begin{minipage}[b]{0.5\textwidth}
    \centering
    \includegraphics[scale=0.4]{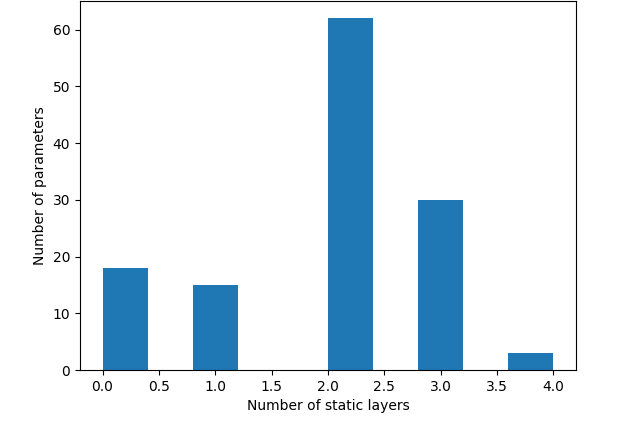}
\end{minipage}%
\begin{minipage}[b]{0.5\textwidth}
    \centering
    \includegraphics[scale=0.4]{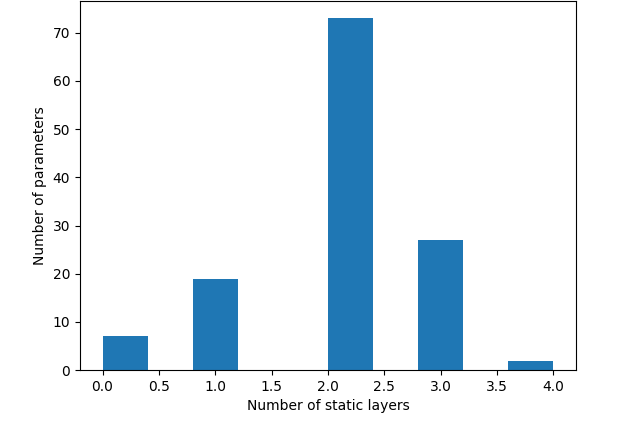}
\end{minipage}
\begin{minipage}[b]{0.5\textwidth}
    \centering
    \includegraphics[scale=0.4]{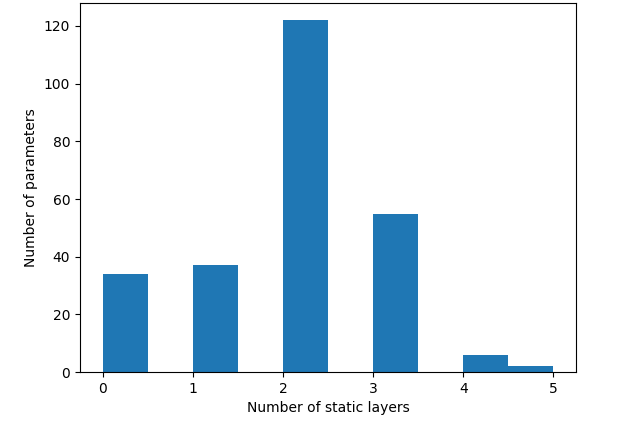}
\end{minipage}%
\begin{minipage}[b]{0.5\textwidth}
    \centering
    \includegraphics[scale=0.4]{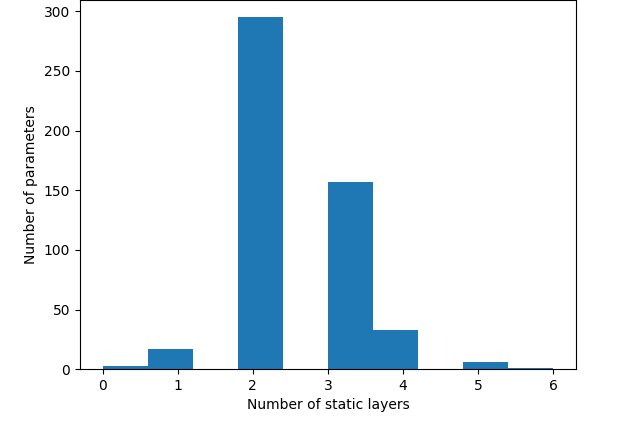}
\end{minipage}
\caption{Number of epochs for which the $\gamma$ parameters have been static, in four different batch normalization layers}
\label{fig:hist_nb_static}
\end{figure}

Depending on the layers, the proportion of static parameters -- as defined in Sec. \ref{staticdynamic} -- varies a lot. While convolutional layers contain mainly dynamic parameters as shown in Fig. \ref{fig:stage2_conv1}, the $\gamma$ parameters in batch normalization layers tend to be mostly static, as can be seen in Fig. \ref{fig:stage2_bn3}. Fig. \ref{fig:hist_nb_static} shows the distribution of the number of epochs for which the $\gamma$ parameters have been static. We can see that most $\gamma$ parameters do not change over the last epoch at least. This explains our definition of static parameters: we seek to select a minimal number of parameters. 

\par 
Considering this, the advantages of batch normalization layers are threefold:
\begin{itemize} \label{batchnormAdvs}
	\item They contain few parameters.
	\item As stated before, $\gamma$ parameters in batch normalization layers influence several input elements and can have a noticeable effect on the following layers.
	\item The large proportion of static parameters means we can protect a few parameters from various batch normalization layers.
\end{itemize} 
\par
Our new strategy is therefore to protect the dynamic parameters from several batch normalization layers. \par 
Since only a few parameters per layer are modified, it is no longer necessary to copy the layer's distribution: making sure the random elements generated are in the range $[0, 0.4]$ is enough to fool a potential attacker who cannot train the model. 

Selecting the static parameters from the four batch normalization layers whose histograms are displayed in Fig. \ref{fig:hist_nb_static} results in protecting 62 parameters (18 in the first layer, 7 in the second, 34 in the fourth and 3 in the last). We replace the selected $\gamma$ parameters by uniformly generated ones in the range $[0, 0.4]$. This leads to the following accuracy:
\begin{itemize}
	\item For $FAR = 10^{-4}$, $FRR = 0.28$ \%
	\item For $FAR = 10^{-5}$, $FRR = 0.86$ \%
\end{itemize}
Even though the drop in the accuracy is less drastic than in the previous experiment on a batch normalizations layer, the $FRR$ for $FAR=10^{-5}$ still corresponds to the accuracy at the end of the 8th training epoch. \par 
Thus, this new method enables us to half the number of parameters to protect while significantly dropping the accuracy (the accuracies obtained are summarized in Table \ref{tab:accuracies}). \par
The question that remains is whether defining static parameters as parameters that have not changed over the last 2 (or more) epochs would lead to an improved security. \par 
Taking now into account the last 2 epochs, and considering 3 batch normalization layers, we have to protect 124 $\gamma$ parameters. This leads to an accuracy of:
\begin{itemize}
	\item For $FAR = 10^{-4}$, $FRR = 0.34$ \%
	\item For $FAR = 10^{-5}$, $FRR = 0.99$ \%
\end{itemize}
Since this new accuracy corresponds to the accuracy at the beginning of the 8th training epoch, we consider that the increased drop in the accuracy does not outweigh the increase in the number of parameters to protect. This confirms our choice of one epoch for the definition of static parameters. \par 

\subsection{Convolutional Layer}\label{convselect}

In this section, we explain how to further drop the accuracy of the degraded modes, while keeping around the same number of protected parameters.

Fig. \ref{fig:conv} shows how a convolutional layer computes the next layer's neurons. A convolutional filter is usually much smaller than the layer's input. Indeed, filters are usually $3 \times 3$ or $5 \times 5$ windows. On the other hand, when dealing with the Labeled Faces in the Wild (LFW) \cite{LFW} dataset, the model's input is commonly 250 $\times 250$ images. Thus, each of the few parameters in a given filter impact a large number of parameters. With the values considered, one filter value modification changes the value of $248 \times 248 = 61,504$ neurons from the following layer. \par 

Therefore, even though convolutional layers have more parameters than batch normalization ones, we can still further limit the number of selected optimal parameters in the convolutional case. \par 

Another element we need to take into account, however, is that the number of filters in a convolutional layer is usually high. For instance, if there are 3 input channels and 64 output channels, the layer stores $64 \times 3 = 192$. Observing any drop in the accuracy requires a change in several such filters. Indeed, feeding degraded values to all the parameters of only two filters among the 192 results in almost no drop in the accuracy. Given the explanation in the previous paragraph, the approach we consider is to randomly select one element among each set of $input\_channels\_number$ filters. Thus, in the previous example, each output channel requires three filters. For each output channel, we randomly select one parameter among the three filters as an optimal parameter. \par 

Furthermore, the depth of the selected convolutional layer matters. Indeed, if the said layer is among the first architecture layers, we can take advantage of the chain reaction. In a convolutional layer, each input neuron impacts several neurons in the following layer due to the way convolutions are computed. Each degraded filter parameter in the considered layer will change the value of a large number of neurons from the following layer, which, in turn, will impact several neurons in the layer after that, and so on. Given that our model is a convolutional neural network, most layers are convolutional ones. This explains why limiting ourselves to few parameters in one convolutional layer at the very beginning of the architecture can lead to a large drop in the accuracy. \par 

Selecting a layer early in the architecture yields three other advantages:
\begin{enumerate}
    \item the number of input channels is lower in the first layers (and only 3 in the first convolutional layer)
    \item the number of output channels is lower in the first layers
    \item the input and output sizes are larger
\end{enumerate}
Points (1) and (2) ensure a minimal overall number of filter parameters for the considered convolutional layer. Point (3) results in a higher impact for every degraded filter parameter. 

Finally, let us note that, given the fact we only consider one parameter per filter, we do not need to take into account the filter's parameters distribution: if the degraded parameters are in the range of possible values, the attacker cannot detect the degradation. 

\begin{figure}
    \centering
    \includegraphics[scale=0.5]{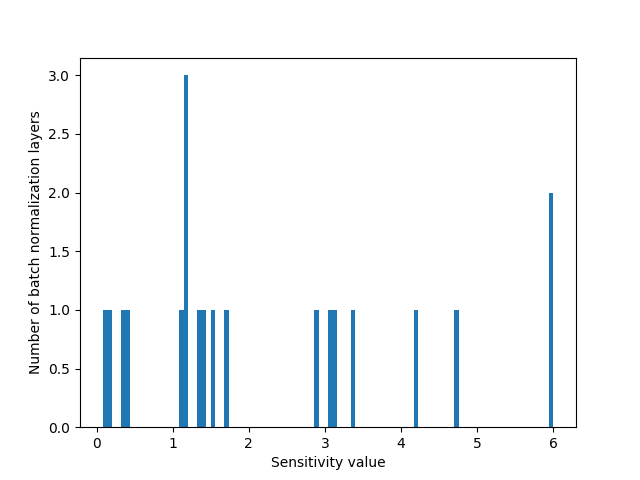}
    \caption{Distribution of the sensitivity with respect to the second image of the CIFAR testing dataset over the batch normalization layers. For each batch normalization layer, the sensitivity $\Delta$ corresponds to the minimum value $\delta$ such that adding $\delta$ to all the $\gamma$ parameters of the layer results in a change in the prediction. All $\Delta$ values greater than 6 are assimilated to 6.}
    \label{batch_pred}
\end{figure}

\begin{figure}
    \centering
    \includegraphics[scale=0.5]{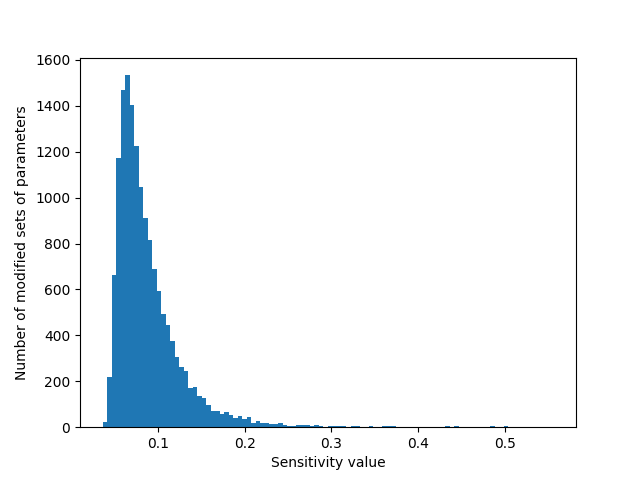}
    \caption{Distribution of the sensitivity with relation to the second image of the CIFAR testing dataset over the first convolutional layer. 64 parameters are selected at random among the layer's parameters, as explained in Sec. \ref{convselect}. For each selected set of parameters, the sensitivity $\Delta$ corresponds to the minimal value $\delta$ such that adding $\delta$ to the selected parameters results in a change in the prediction.}
    \label{sens_neurons}
\end{figure}

To summarize, the strategy to select the optimal parameters is as follows:
\begin{itemize}
    \item Consider the model's first convolutional layer
    \item For each output channel, select one element among the three filters for that channel
\end{itemize}
In order to check that the first convolution has a higher impact on the predictions than batch normalization layers, we compare the sensitivity (as explained in \ref{minparams}) of the two strategies on a ResNet18 architecture trained on the CIFAR10 dataset \cite{cifar}. Thus, we select one image (the second image from the CIFAR10 testing set for instance), and plot, on the one hand, the minimum $\delta$ one needs to add to all the $\gamma$ parameters of each batch normalization layer in order to change the model's prediction (Fig.\ref{batch_pred}), and, on the other hand, the minimum $\delta$ to add to 64 parameters from the first convolutional layer, randomly selected according to our strategy (Fig. \ref{sens_neurons}). These figures confirm that for a given input, the sensitivity of $\gamma$ parameters in batch normalization layers is lower than that of the convolutional parameters, selected according to our strategy.   

For our model, this strategy results in 64 selected parameters. As before, we can encode each parameter on 8 bits, thus leading to all the parameters being encoded on 512 bits overall. 
When we change 64 parameters from the model's first convolutional layer -- selected as described previously -- to random ones, we get the following accuracy:
\begin{itemize}
	\item For $FAR = 10^{-4}$, $FRR = 0.34$ \%
	\item For $FAR = 10^{-5}$, $FRR = 0.97$ \%
\end{itemize}
Thus, with only half the parameters, we reach almost the same in accuracy as the batch normalization layer's case. 
\begin{table}[ht]
\caption{Accuracy of the original model and the of the model where some parameters have been replaced by random ones} \label{tab:accuracies}
\begin{tabular}{m{4.5cm} m{2.5cm} m{2.5cm} m{2.5cm}}  
\toprule
                                            & Number of protected parameters    & $FRR$ ($FAR=10^{-4})$ & $FRR$ ($FAR=10^{-5})$ \\
                                            \midrule \midrule
 Original model                             & 0     & 0.24 \%                   & 0.70 \% \\ \midrule
                                                                   
 Modification of one batch normalization (beginning of Sec. \ref{batchnormSec})   & 128   & 0.32  \%                  & 0.96 \% \\ \midrule
 Modification of dynamic parameters (end of Sec. \ref{batchnormSec})        & 62    & 0.28 \%                   & 0.86 \% \\ \midrule
 Modification of convolutional parameters (Sec. \ref{convselect})   & 64    & 0.34 \%                   & 0.97 \% \\ 
 
 \bottomrule
\end{tabular} 
\end{table}

\subsection{Exhaustive search FAR and FRR} \label{es}
We place ourselves in the attacker's shoes. Thus, we generate random 16 digit PINs and compute the accuracy -- FRR for FAR -- associated with the deduced parameters instead of the optimal ones.

The minimal accuracy the attacker gets is the following for degraded parameters from the first convolutional layer (selected as in Sec. \ref{convselect}):
\begin{itemize}
	\item For $FAR = 10^{-4}$, $FRR = 0.30$ \%, representing a 25\% relative increase compared to the original model
	\item For $FAR = 10^{-5}$, $FRR = 0.94$ \%, representing a 34\% relative increase compared to the original model
\end{itemize}

On average, the attacker gets:
\begin{itemize}
	\item For $FAR = 10^{-4}$, $FRR = 0.87$ \%, representing a 263\% relative increase compared to the original model
	\item For $FAR = 10^{-5}$, $FRR = 2.58$ \%, representing a 269\% relative increase compared to the original model
\end{itemize}

For the second strategy on the batch normalization layers (Sec. \ref{batchnormSec}), the attacker gets, on average, the following accuracies:
\begin{itemize}
	\item For $FAR = 10^{-4}$, $FRR = 0.28$ \%, representing a 16\% relative increase compared to the original model
	\item For $FAR = 10^{-5}$, $FRR = 0.82$ \%, representing a 17\% relative increase compared to the original model
\end{itemize}

To gauge the accuracy of our system, we use again the LFW database and a proprietary setup. Each try takes around 15 minutes. 

\section{Conclusion} \label{Conclusion}

We introduce a premium mode for NN applications in mobile phones.
Our defense strategy is threefold: 
\begin{itemize}
    \item we rely on a PIN only known by privileged users;
    \item the functionality of the NN is degraded by default;
    \item the attacker does not have access to a training dataset and a limited testing facility and is therefore forced to blindly guess the correct PIN.
\end{itemize}

Each privileged user benefits from a dedicated training of the NN and is given a PIN which enables him to switch from a degraded mode to the premium one. \par 
These protections can also be enforced by classical anti-reversing engineering techniques as well as OS and software security features. \par
We explain how for a facial recognition NN with more than 14 million parameters, we determine 64 sensitive optimal values for our proposal, showing its practicability.  \par 
We think that we can extend our application domain 
to health or marketing ones, where data sets are also difficult to come by.

\subsubsection{Acknowledgments} The authors want to thank Vincent Despiegel and his team.

\bibliographystyle{splncs04}
\bibliography{bibli}
\end{document}